\documentclass[a4paper,preprintnumbers,showpacs,twocolumn,superscriptaddress,nofootinbib,amsmath,amssymb]{revtex4-1}
\usepackage[dvips]{graphics}
\usepackage[hypertex]{hyperref}
\usepackage{color,hyperref}
\usepackage{times}
\usepackage{mathrsfs}
\hypersetup{
  colorlinks=true,
  linkcolor=blue,
  citecolor=magenta,
  filecolor=magenta,
  urlcolor=blue
}

\def\beq{\begin{equation}}
\def\eeq{\end{equation}}
\def\bear{\begin{eqnarray}}
\def\ear{\end{eqnarray}}
\def\nn{\nonumber}

\usepackage{enumitem}
\usepackage{graphicx,subfigure}% Include figure files
\usepackage{dcolumn}% Align table columns on decimal point
\usepackage{bm}% bold math
\usepackage{color}

\begin{document}

\title{Spinning test particles in the $\gamma$ space-time}

\author{Bobir Toshmatov}
\email{bobir.toshmatov@nu.edu.kz} \affiliation{Department of
Physics, Nazarbayev  University, 53 Kabanbay Batyr, 010000
Nur-Sultan, Kazakhstan} \affiliation{Ulugh Beg Astronomical
Institute, Astronomicheskaya 33, Tashkent 100052, Uzbekistan}

\author{Daniele Malafarina}
\email{daniele.malafarina@nu.edu.kz} \affiliation{Department of
Physics,  Nazarbayev University, 53 Kabanbay Batyr, 010000
Nur-Sultan, Kazakhstan}

\begin{abstract}

We consider the motion of spinning particles in the field of a
well known vacuum static axially-symmetric space-time, known as
$\gamma$-metric, that can be interpreted as a generalization of
the Schwarzschild manifold to include prolate or oblate
deformations. We derive the equations of motion for spinning test
particles by using the Mathisson-Papapetrou-Dixon equations
together with the Tulczyjew spin-supplementary condition, and
restricting the motion to the equatorial plane. We determine the
limit imposed by super-luminal velocity for the spin of the
particle located at the innermost stable circular orbits (ISCO).
We show that the particles on ISCO of the prolate $\gamma$
space-time are allowed to have higher spin than the corresponding
ones in the the oblate case. We determine the value of the ISCO
radius depending on the signature of the spin-angular momentum,
${\rm s-L}$ relation, and show that the value of the ISCO with
respect to the non spinning case is bigger for ${\rm sL}<0$ and
smaller for ${\rm sL}>0$. The results may be relevant for
determining the properties of accretion disks and constraining the
allowed values of quadrupole moments of astrophysical black hole
candidates.

\end{abstract}

\maketitle

\section{Introduction}\label{sec-intr}

Most of the current astrophysical observations of extreme compact
objects such as black hole candidates, are obtained from light
emitted by matter accreting around the compact object. It is
natural to assume that matter in the accretion disks moves along
geodesics and therefore from the study of geodesics we can infer
useful information on the background geometry of the central
object. In particular, circular orbits that are located close to
the infinitely redshifted surface are extremely useful, as they
provide information about the strong field regime and possibly the
nature of the central object itself. Two of the most important
circular orbits around compact objects are the light ring (photon
sphere) and the innermost stable circular orbit (ISCO) for massive
particles. In the static and spherically symmetric case, i.e. in
the Schwarzschild case, the light ring and ISCO are located at
$r=3M$ and $r=6M$, respectively~\cite{Felice:1968}. If the case of
rotating black holes, i.e. in the Kerr case, the scenario is more
complicated as locations of the characteristic orbits depend on
the direction of the particle's orbital angular momentum ($\rm L$)
and the spin of the central object ($a$). More precisely, if the
particle is co-rotating, i.e., moving in the rotation direction of
the central object ($a{\rm L}>0$), then the radii of the light
ring and ISCO decrease with respect to the Schwarzschild case and
at the extreme value of the rotation parameter ($a=M$) they
coincide at $r=M$. On the other hand, if the particle is
counter-rotating, i.e., moving in the opposite rotation direction
with respect to the central object ($a{\rm L}<0$), then the radii
of the light ring and ISCO increase with respect to the
Schwarzschild case, and at the extreme value of the rotation
parameter they become $r=4M$ and $r=9M$,
respectively~\cite{Bardeen:APJ:1972,Pugliese:PRD:2011,disk}.
Similar results are obtained in the Kerr-Newmann family of
space-times if one includes charge~\cite{Carter:PR:1968}.

The recent detection of gravitational waves by LIGO and VIRGO have
confirmed that the black holes~\cite{GW150914} and neutron
stars~\cite{GW170817} in the coalescence of the binary system are
spinning. Indeed, in these events the masses of the two objects in
the binary system were comparable. However, there exist
astrophysical scenarios where one component of the binary system
has negligible mass, $m$, as compared to the companion with mass
$M$ (i.e. as $m\ll M$), thus making the test particle
approximation a valid tool to determine the characteristic orbit.
Typically, the case of dust particles orbiting a stellar mass
black hole and the case of a neutron star orbiting a supermassive
black hole fit in the above description.

The motion of spinning test particle in non-homogeneous
gravitational fields has been considered in several articles (see
\cite{Barausse:PRD:2009,Steinhoff:PRD:2012,LG:PRD:2016a,LG:PRD:2014}
and references therein). Most of these studies are restricted to
the ``pole-dipole" approximation where just monopole (mass) and
dipole (rotational angular momentum, i.e., spin) are taken into
account~\cite{Steinhoff:PRD:2010}. The equations of motion of such
systems are described by the Mathisson-Papapetrou-Dixon (MPD)
equations~\cite{Mathisson:1937,PP:1951,Dixon:1970} with some
spin-supplementary condition (SSC). The validity of the MPD
equations in the limit of strong fields has been discussed
in~\cite{Ramirez:2017pmp,Deriglazov:2017jub}. The SSC serves as a
reference point inside the spinning body whose evolution is
described by the equations of motion. In the literature, several
SSCs have been proposed, such as, Tulczyjew \cite{Tulczyjew:1959},
Pirani \cite{Pirani:1956}, etc. -- for details, see
\cite{LG:PRD:2014}. Similarly, the characteristic orbits of
spinning particles in non-rotating and rotating axially symmetric
space-times, are shifted inward or outward depending on signature
of the spin of the particle, with respect to the non-spinning
case~\cite{Hojman:PRD:1977,Abramowicz:MNRAS:1979,Suzuki:PRD:1998,Stuchlik:APS,Semerak:MNRAS:1999,Stuchlik:CQG,Semerak:MNRAS:2007,Plyatsko:PRD:2013,Han:2008zzf,Han:2010tp,Hackmann:PRD:2014,Jefremov:PRD:2015,Zhang:PRD:2018}.

In this paper, by using the ``pole-dipole" approximation, we study
the motion of spinning particles in the $\gamma$-metric. The
$\gamma$ metric, also known as Zipoy-Vorhees space-time, is an
asymptotically flat, vacuum solution of Einstein's equations which
belongs to the Weyl class of static, axially symmetric
space-times~\cite{Zip,Vor}. The $\gamma$ metric is fully
characterized by only two parameters: one, $M$, related to the
mass of the source, and the other, $\gamma$, which can be called
deformation parameter, related to the shape of the source. The
metric is continuously linked to the Schwarzschild metric through
the value of $\gamma$, as the spherically symmetric case is
recovered for $\gamma=1$. The cases $0<\gamma<1$ and $\gamma>1$
represent sources with prolate and oblate spheroidal deformations,
respectively. It is important to stress that in these cases, i.e.
for $\gamma \neq 1$, the line element does not represent a black
hole~\cite{Papadopoulos:PRD:1981,Bonnor:GRG,Herrera:IJMPD}, as the
surface $r=2M$ becomes a true curvature singularity. The motion of
test particles in the $\gamma$ metric was considered in
\cite{herrera1,Boshkayev:PRD:2016,us1,Hernandez-Pastora:PRD,Abdikamalov:PRD:2019,Toshmatov:PRD:2019}
and it was shown that the space-time can be considered as black
hole ``mimicker" and constitutes an excellent candidate to study
possible astrophysical tests of black hole candidates.

The paper is organized as follows: in Sec.
\ref{sec-spinning-particle}, we present the general formalism for
spinning particles in the ``pole-dipole" approximation, i.e., MPD
equations with Tulczyjew - SSC and, we derive the equations of
motion for spinning particles in a generic space-time. In Sec.
\ref{sec-gamma-metric} we apply derived equations to the
$\gamma$-metric and calculate the ISCO for spinning particles and
compare it with the Schwarzschild case. Finally, in Sec.
\ref{sec-conclusion} we summarize the results and discuss how they
could be relevant for astrophysical observations of black hole
candidates. Throughout the paper, we use natural units setting
$G=c=1$.

\section{Dynamics of spinning particles}\label{sec-spinning-particle}

The equations of motion of spinning test particles are determined by the
MPD equations, which can be given as~\cite{Mathisson:1937,PP:1951,Dixon:1970}
\bear &&\frac{Dp^\alpha}{d\lambda}=-\frac{1}{2} R^{\alpha}
_{\beta\delta\sigma}u^\beta S^{\delta\sigma}\ ,\label{MPD-eq1}\\
&&\frac{DS^{\alpha\beta}}{d\lambda}=p^\alpha u^\beta-u^\alpha
p^\beta\ ,\label{MPD-eq2} \ear where $D/d\lambda$ is the covariant
derivative along the particle's trajectory ($D/d\lambda\equiv
u^\alpha\nabla_\alpha$), $\lambda$ is the affine parameter,
$R^{\alpha} _{\beta\delta\sigma}$ is the Riemann tensor,
$p^\alpha$ and $u^\alpha$ are the dynamical 4-momentum and
kinematical 4-velocity of the particle, respectively, and
$S^{\alpha\beta}$ is the spin tensor. Notice that
$S^{\alpha\beta}$ is anti-symmetric (i.e.
$S^{\alpha\beta}=-S^{\beta\alpha}$) and so it has only six
independent components. Obviously, the spinning particle does not
follow a geodesic trajectory because of the spin-curvature force
$R^{\alpha} _{\beta\delta\sigma}u^\beta S^{\delta\sigma}$.

To solve equations~(\ref{MPD-eq1}) and (\ref{MPD-eq2}), we need
one extra condition. Therefore, in order to restrict the spin
tensor to generate rotations only, we employ the so called
``Tulczyjew spin-supplementary condition"
(SSC)~\cite{Tulczyjew:1959} given by
\bear\label{SSC-Tulcz}
S^{\alpha\beta}p_{\alpha}=0\ .
\ear
Then, from the above SSC, it turns out that both the canonical momentum
and the spin of the particle are conserved quantities as
\bear
&&p^{\alpha}p_{\alpha}=-m^2\ ,\label{mass-conservation}\\
&&S^{\alpha\beta}S_{\alpha\beta}=2S^2\ .\label{spin-conservation}
\ear
However, it is worth noticing that, despite the canonical momentum
of the spinning particle being conserved, its squared velocity
does not necessarily satisfy the normalization condition $u_\alpha
u^\alpha=-1$, as the 4-vectors $p^\alpha$ and $u^\alpha$ are not
always parallel. Furthermore, in addition to the SSC-dependent
conserved quantities, there are the usual background-dependent
conserved quantities associated with the Killing vectors,
$\xi^\alpha$, which can be expressed as
\bear\label{killing-conservation}
C_{\xi}=p^\alpha\xi_\alpha-
\frac{1}{2}S^{\alpha\beta}\nabla_\beta\xi_\alpha\ .
\ear
Since the spin tensor is antisymmetric and the Christoffel
symbols $\Gamma^\nu_{\alpha\beta}$ are symmetric, it immediately
follows that $S^{\alpha\beta}\Gamma^\nu_{\alpha\beta}=0$, and
consequently, the background-dependent conserved quantities via
Killing vectors~(\ref{killing-conservation}) can be written as
\bear\label{killing-conservation2} C_{\xi}=p^\alpha\xi_\alpha-
\frac{1}{2}S^{\alpha\beta}\partial_\beta\xi_\alpha\ . \ear

The line element of a generic stationary axially symmetric space-time is
given by
\bear\label{line-element}
ds^2=g_{tt}dt^2+g_{rr}dr^2+2g_{t\phi}dtd\phi
+g_{\theta\theta}d\theta^2+g_{\phi\phi}d\phi^2\ ,
\ear
where the metric functions depend on the coordinates $r$ and
$\theta$. From the symmetry of the metric one can easily see that
the line element~(\ref{line-element}) allows for two Killing
vector fields, one related to time translations and one to
rotations, as given by
\bear
\xi^\alpha=\delta^\alpha_t, \qquad
\xi^\alpha=\delta^\alpha_\phi.
\ear
The corresponding conserved quantities, i.e. energy and angular momentum, can be written as
\bear
&&{\rm E}=-p_t+\frac{1}{2}g_{t\alpha,\beta}S^{\alpha\beta},\label{energy}\\
&&{\rm L}=p_\phi+\frac{1}{2}g_{\phi\alpha,\beta}S^{\alpha\beta}.\label{angular-momentum}
\ear
When considering astrophysical applications, such as accretion
disks, it is sufficient to consider test particles moving on the
equatorial plane, $\theta=\pi/2$.  When restricted to the
equatorial plane the metric functions depend only on the radial
coordinate and $p^\theta=0$. The number of independent components
of the spin tensor is reduced to three since
\bear\label{spin-theta}
S^{\theta\alpha}=0\ ,
\ear
and for the remaining components, from the
Tulczyjew-SSC~(\ref{SSC-Tulcz}), one finds the following
relations:
\bear &&S^{t\phi}=-\frac{p_r}{p_\phi}S^{tr}\ ,
\label{SSC-betat}\\
&&S^{r\phi}=\frac{p_t}{p_\phi}S^{tr}\ . \label{SSC-betar}
\ear
From the normalization condition~(\ref{mass-conservation}) one
finds the radial momentum of the particle given by
\bear\label{momentum-r}
p_r^2=g_{rr}\left(-g^{tt}p_t^2-g^{\phi\phi}p_\phi^2-2g^{t\phi}p_tp_\phi-m^2\right)\ .
\ear
By using the relations~(\ref{SSC-betat}), (\ref{SSC-betar}), and
(\ref{momentum-r}) from the spin conservation
law~(\ref{spin-conservation}), one finds the $(t,r)$-component of
the spin tensor as
\bear\label{Str}
S^{tr}=\pm\frac{p_\phi
s}{\sqrt{g_{rr}(g_{t\phi}^2-g_{tt}g_{\phi\phi})}}
\ear
where $\pm$ signs represent the direction of spin with respect to
direction of $p_\phi$. In expression (\ref{Str}) we have written
the spin parameter $S$ in terms of the the specific spin angular
momentum of the particle $s$ as $S=ms$.
%
%\bear
%S=ms\ .
%\ear
%
Finally, from the conservation of energy~(\ref{energy}) and
angular momentum~(\ref{angular-momentum}), we find the $t$ and
$\phi$ components of the four-momentum as
\bear
&&p_t=\frac{{\rm -E+s(AL+BE)}}{1-s^2{\rm D}}\ ,\label{energy1}\\
&&p_\phi=\frac{{\rm L+s(BL+CE)}}{1-s^2{\rm D}}\ .\label{angular-momentum1}
\ear
with
\bear
&&{\rm A}=\frac{g_{tt}'}{2\sqrt{g_{rr}(g_{t\phi}^2-g_{tt}g_{\phi\phi})}},\nonumber\\
&&{\rm B}=\frac{g_{t\phi}'}{2\sqrt{g_{rr}(g_{t\phi}^2-g_{tt}g_{\phi\phi})}},\nonumber\\
&&{\rm C}=\frac{g_{\phi\phi}'}{2\sqrt{g_{rr}(g_{t\phi}^2-g_{tt}g_{\phi\phi})}},\nonumber\\
&&{\rm
D}=B^2-AC=\frac{(g_{t\phi}')^2-g_{tt}'g_{\phi\phi}'}{4g_{rr}
\left(g_{t\phi}^2-g_{tt}g_{\phi\phi}\right)},\nonumber
\ear where
prime denotes the partial derivative with respect to radial
coordinate as $f'\equiv\partial f/\partial r$. The contravariant
forms of the momenta are determined by
$p^\alpha=g^{\alpha\beta}p_\beta$.

By substituting
expressions~(\ref{energy1}) and~(\ref{angular-momentum1}) into the
radial component of the four-momentum~(\ref{momentum-r}) one
arrives at the expression
\bear\label{momentum-r2}
\left(p^r\right)^2=\frac{\beta}{\alpha}({\rm E}-V_+)({\rm E}-V_-)\
,
\ear
where the effective potentials $V_\pm$ are given by
\bear\label{eff-potential}
V_{\pm}=-\frac{\delta {\rm
L}}{\beta}\pm\sqrt{\frac{\delta^2{\rm
L}^2}{\beta^2}+\frac{\rho-\sigma {\rm L}^2}{\beta}}\ ,
\ear
and
\begin{widetext}
\bear\label{notations1}
\alpha&=&g_{rr}\left[1-\frac{s^2\left((g_{t\phi}')^2-g_{tt}'g_{\phi\phi}'\right)}{4g_{rr}
\left(g_{t\phi}^2-g_{tt}g_{\phi\phi}\right)}\right]^{2}\ ,\nonumber\\
\beta&=&-g^{tt}+\frac{s\left(g^{tt}g_{t\phi}'+g^{t\phi}
g_{\phi\phi}'\right)}{\sqrt{g_{rr}(g_{t\phi}^2-g_{tt}g_{\phi\phi})}}
%+ \nonumber \\&&
-\frac{s^2
\left[g^{tt}(g_{t\phi}')^2+g_{\phi\phi}'\left(2g^{t\phi}
g_{t\phi}'+g^{\phi\phi} g_{\phi\phi}'\right)\right]}{4
g_{rr}\left(g_{t\phi}^2-g_{tt}g_{\phi \phi}\right)}\ ,\nonumber\\
\delta&=&g^{t\phi}+\frac{s\left(g^{tt}g_{tt}'-g^{\phi\phi}
g_{\phi\phi}'\right)}{2\sqrt{g_{rr}(g_{t\phi}^2-g_{tt}g_{\phi\phi})}}
%+\\&&
-\frac{s^2
\left[g_{t\phi}'\left(g^{tt}g_{tt}'+g^{t\phi}g_{t\phi}'\right)+g_{\phi\phi}'\left(g^{t\phi}
g_{tt}+g^{\phi\phi}g_{t\phi}'\right)\right]}{4 g_{rr}\left(g_{t\phi}^2-g_{tt}g_{\phi\phi}\right)}\ ,\nonumber \\
\sigma&=&-g^{\phi\phi}-\frac{s\left(g^{t\phi}g_{tt}'-g^{\phi\phi}
g_{t\phi}'\right)}{\sqrt{g_{rr}(g_{t\phi}^2-g_{tt}g_{\phi\phi})}}
%+ \nonumber \\ &&
-\frac{s^2
\left[g^{tt}(g_{tt}')^2+g_{t\phi}'\left(2g^{t\phi}g_{tt}'+g^{\phi
\phi}g_{t\phi}'\right)\right]}{4 g_{rr}\left(g_{t\phi}^2-g_{tt}g_{\phi\phi}\right)}\ ,\nonumber\\
\rho&=&m^2\left[1-\frac{s^2\left((g_{t\phi}')^2-g_{tt}'g_{\phi\phi}'\right)}{4g_{rr}
\left(g_{t\phi}^2-g_{tt}g_{\phi\phi}\right)}\right]^2\
.\nonumber
\ear
\end{widetext}

One can see from (\ref{momentum-r2}) that in order to have
$(p^r)^2\geq0$, the energy of the particle must satisfy one of the
following conditions:
\bear
&&{\rm E}\in(-\infty, V_-],\label{condition1}\\
&&{\rm E}\in[V_+,\infty).\label{condition1}
\ear
In the following, we will restrict our attention to the case of
test particles with positive energy and thus will restrict the
attention to the effective potential $V_{\rm eff}=V_+$.

Let us focus on the characteristic circular orbits of the spinning
test particle in the space-time described by the line element (\ref{line-element}). It is
well known that circular motion of particles moving in the
central field is governed by the following conditions:
\begin{itemize}
\item[(i)] The radial velocity of the particle must vanish at the circular orbit. Namely
\bear\label{cond-circular1}
\frac{dr}{d\lambda}=0, \quad \text{which implies} \quad V_+=E\ ,
\ear
\item[(ii)] The radial acceleration of the particle must
vanish. Namely
\bear\label{cond-circular2}
\frac{d^2r}{d\lambda^2}=0, \quad \text{which implies} \quad \frac{dV_+}{dr}=0\ .
\ear
\end{itemize}
However, these conditions do not in general guarantee that the
circular orbits are stable. Stability of the orbit is provided by
positivity of the second derivative of the effective potential
with respect to radial coordinate as
\bear\label{cond-isco}
\frac{d^2V_+}{d\lambda^2}\geq0\ ,
\ear
with the equality holding
for the marginally stable orbits, corresponding to the smallest
allowed value for stable circular orbits, namely the ISCO.

Before proceeding to the motion of spinning test particles in the
$\gamma$-metric, there is one more important feature which appears
due to the spin of the particle that should be considered. That is
the super-luminal bound on the particle's motion. As it was
mentioned before, the dynamical 4-momentum and kinematical
4-velocity of the spinning particle are not always parallel.
Therefore, the normalization $u_\alpha u^\alpha=-1$ does not hold
while, $p_\alpha p^\alpha=-m^2$ is satisfied. As the spinning
particle approaches the center of the space-time, its 4-velocity
increases and eventually, for certain values of the spin and
radius some components of the 4-velocity may diverge as $u_\alpha
u^\alpha\rightarrow+\infty$. Before this to happen, the particle's
motion crosses the boundary between time-like and space-like
trajectories. Of course, space-like (i.e. super-luminal) motion is
physically meaningless, and the transition to $u_\alpha
u^\alpha>0$ is not allowed for real particles. Therefore one must
impose a further bound, called super-luminal bound, which is
defined by the relation $u_\alpha u^\alpha=0$.

Within the Tulczyjew-SSC (\ref{SSC-Tulcz}), one can find the
components of the 4-velocity $u^\alpha$ from the following
velocity-momentum relation~\cite{Kunzle:JMP:1972}:
\bear\label{velocity-momentum}
u^\alpha=\frac{\mu}{m^2}\left(p^\alpha+\frac{2S^{\alpha\beta}R_{\beta
\delta \sigma \rho }p^{\delta}S^{\sigma\rho }}{4m^2+R_{abcd}S^{ab}
S^{cd}}\right)\ ,
\ear
where $\mu$ is kinematical mass (or rest mass) of the particle and
it is defined by $u^\alpha p_\alpha=-\mu$. Explicit analytical
forms of components of the 4-velocity, $u^\alpha$, in a generic
stationary space-time~(\ref{line-element}) are very long. However,
in the static case (i.e. $g_{t\phi}=0$), for motion restricted to
the equatorial plane, they reduce to

\begin{widetext}
\bear\label{ualpha}
&&u^t=p_t\left\{g^{tt}+\frac{\left(S^{tr}\right)^2}{X}\left[
\frac{p_r^2}{p_\phi^2}\left(\frac{B_1}{g_{rr}}+\frac{C_1}{g_{tt}}\right)
-\frac{A_1}{g_{tt}}+\frac{B_1}{g_{\phi\phi}}\right]\right\}\ ,\nonumber\\
&&u^r=p_r\left\{g^{rr}+\frac{\left(S^{tr}\right)^2}{X}\left[
-\frac{p_t^2}{p_\phi^2}\left(\frac{B_1}{g_{rr}}+\frac{C_1}{g_{tt}}\right)
+\frac{A_1}{g_{rr}}+\frac{C_1}{g_{\phi\phi}}\right]\right\}\ ,\nonumber\\
&&u^\theta=0\ ,\\
&&u^\phi=p_\phi\left\{g^{\phi\phi}+\frac{\left(S^{tr}\right)^2}{X}\left[
\frac{p_t^2}{p_\phi^2}\left(\frac{A_1}{g_{tt}}-\frac{B_1}{g_{\phi\phi}}\right)
+\frac{p_r^2}{p_\phi^2}\left(\frac{A_1}{g_{rr}}+
\frac{C_1}{g_{\phi\phi}}\right)\right]\right\}\ ,\nonumber
\ear
\end{widetext}
where
\bear\label{coeffs}
&&X=4m^2+\left(S^{tr}\right)^2\left(-A_1+\frac{p_t^2}{p_\phi^2}B_1
+\frac{p_r^2}{p_\phi^2}C_1\right)\
,\nonumber\\
&&A_1=\frac{g_{rr}'g_{tt}'}{g_{rr}}+\frac{(g_{tt}')^2}{g_{tt}}-2g_{tt}''\
,\nonumber\\
&&B_1=\frac{g_{rr}'g_{\phi\phi}'}{g_{rr}}+\frac{(g_{\phi\phi}')^2}{g_{\phi\phi}}
-2g_{\phi\phi}''\ ,\\
&&C_1=\frac{g_{tt}'g_{\phi\phi}'}{g_{rr}}\ .\nonumber
\ear
Since in this paper, we are mainly concerned with circular orbits
for the spinning test particle, and in particular with the ISCO,
let us explicitly evaluate the super-luminal limit of spinning
particles on circular orbits. In the case of circular orbits, the
relation $u^\alpha u_\alpha=0$ becomes:
\bear\label{super-generic}
\frac{g_{tt}g_{\phi\phi}p_t^2p_\phi^2\left(S^{tr}
\right)^4}{\left[4p_\phi^2-\left(S^{tr}\right)^2\left(B_1p_t^2
-A_1p_\phi^2\right)\right]^2}
\left(\frac{B_1}{g_{\phi\phi}}-\frac{A_1}{g_{tt}}\right)^2=1\
,\nonumber\\
\ear
Notice that in (\ref{super-generic}) the momentum and spin tensors
are functions of the spin and therefore, given the non trivial
dependence on $s$, equation (\ref{super-generic}) cannot be solved
analytically, even in the simplest case that is the Schwarzschild
space-time.

Alternatively, one can use the method developed
in~\cite{Hojman:CQG:2013}. From the second of the MPD
equations~(\ref{MPD-eq2}) and the Tulczyjew-SSC (\ref{SSC-Tulcz})
one finds the following relations~\footnote{By following the gauge
choices and invariant relations in \cite{Hojman:CQG:2013}, the
notation $\lambda=t$ is adopted.}:
\bear
\frac{DS^{tr}}{d\lambda}&&=-\frac{p_\phi}{p_t}\frac{DS^{\phi r}}{d\lambda}-
\frac{S^{\phi r}}{p_t}\frac{Dp_\phi}{d\lambda}+
S^{\phi r}\frac{p_\phi}{p_t^2}\frac{Dp_t}{d\lambda}\nonumber\\
&&=p^tu^r-p^r\ ,\label{str-eq}\\
\frac{DS^{t\phi}}{d\lambda}&&=\frac{p_r}{p_t}\frac{DS^{\phi r}}{d\lambda}+
\frac{S^{\phi r}}{p_t}\frac{Dp_r}{d\lambda}-
S^{\phi r}\frac{p_r}{p_t^2}\frac{Dp_t}{d\lambda}\nonumber\\
&&=p^tu^\phi-p^\phi\ ,\label{stp-eq}
\ear
where
\bear
&&\frac{DS^{\phi r}}{d\lambda}=p^\phi u^r-u^\phi p^r\ ,\\
&&S^{\phi r}=\mp\frac{p_t
s}{\sqrt{g_{rr}(g_{t\phi}^2-g_{tt}g_{\phi\phi})}}\ ,
\ear
and from the first of the MPD equations~(\ref{MPD-eq1}), one can write
the covariant derivatives of the components of the dynamical
4-momentum as
\bear\label{Dpalpha}
\frac{Dp_\alpha}{d\lambda}=-\frac{1}{2}R_{\alpha\beta\delta\sigma}u^\beta S^{\delta\sigma}\ .
\ear
By using the above relations, one can solve equations
(\ref{str-eq}) and (\ref{stp-eq}) with respect to $u^r$ and
$u^\phi$, simultaneously. Thus,
for the particle to move always in the time-like region,
we must impose the following condition:
\bear\label{super-luminal}
\frac{u_\alpha
u^\alpha}{(u^t)^2}=g_{tt}+g_{rr}(u^r)^2+2g_{t\phi}u^\phi
+g_{\phi\phi}(u^\phi)^2\leq0\ ,
\ear
with equality holding for the super-luminal bound. In the next
section we will adapt the above formalism to a background
space-time given by the $\gamma$-metric and determine how the
particle's spin in this space-time affects the motion in
comparison with the Schwarzschild and Kerr space-times.

%%%%%%%%%%%%%%%%%%%%%%%%%%%%%%%%%%%%%%%%%%%%%%%%%%%%%%%%%%%%%%%%%%%%

\section{Spinning particle in the $\gamma$-metric}\label{sec-gamma-metric}

The $\gamma$ space-time, is a static axially symmetric vacuum
solution of Einstein's equations that is represented by the line
element~\cite{Zip,Vor}
\bear\label{line-element-gamma}
ds^2=-f^\gamma
dt^2+&&f^{\gamma^2-\gamma}g^{1-\gamma^2} \left(\frac{dr^2}{f}
+r^2d\theta^2\right)+\nonumber\\&&+f^{1-\gamma}r^2\sin^2\theta
d\phi^2\ ,
\ear
where
\bear
&&f(r)=1-\frac{2M}{r}\ ,\\
&&g(r,\theta)=1-\frac{2M}{r}+\frac{M^2\sin^2\theta}{r^2}\ .\nonumber
\ear
The line-element depends on two parameters, $M$, related to the
mass of the source and $\gamma$, related to its deformation from
spherical symmetry. To understand the meaning of such parameters
we may consider the asymptotic expansion in multipoles of the
gravitational potential~\cite{hernandez}. Then it is easy to see
that the total mass of the source (i.e. the monopole moment) as
measured by an observer at infinity is $M_{\rm tot}=M\gamma$ and
the quadrupole moment  is $Q=M^3\gamma(1-\gamma^2)/3$. The
striking difference between the case $\gamma=1$ (i.e.
Schwarzschild) and the case with non vanishing quadrupole moment
comes from the analysis of the Kretschmann scalar which shows that
the surface $r=2M$ is a curvature singularity~\cite{virb}.
Therefore the space-time is geodesically incomplete with the
radial coordinate limited to values $r\in(2M,\infty)$. One
traditional interpretation of the $\gamma$-metric suggests that
static compact objects must tend to become spherical as they
become more compact to the limit that they must shed away all
higher multipole moments as they cross the horizon at $r=2M$.
However, perfect spherical symmetry is a mathematical abstraction
that is not expected to exist in the real world. Therefore,
another possible interpretation of the $\gamma$-metric can be
considered if we are to understand the singularity as a regime
where classical relativistic description fails. Then, we may allow
for the existence of exotic compact objects with non vanishing
quadrupole moment and we can interpret the surface $r=2M$ as the
boundary of such object. The region close to $r=2M$ would then
need a theory of quantum-gravity to be described and its high
red-shift (that tends to infinity in the classical limit) would
make such objects look like black holes to distant observers. We
now ask the question whether, in principle, we can be able to
distinguish a black hole space-time from the $\gamma$-metric by
observing the motion of spinning particles in accretion disks.

From (\ref{Str}) one can write $(t,r)$ component of the spin
tensor as
\bear\label{Str-gamma}
S^{tr}=\frac{sp_\phi}{r}\left(1-\frac{2M}{r}\right)^{(1-
\gamma)\gamma/2}\left(1-\frac{M}{r}\right)^{-1+\gamma^2}\ ,
 \ear

Then, from~(\ref{energy1}) and (\ref{angular-momentum1})
one can write the covariant momenta of the neutral spinning
test particle corresponding to the $t$ and $\phi$ coordinates as
\bear
&&p_t=-\frac{{\rm E+sAL}}{1-s^2{\rm D}}\ ,\label{pt-gamma}\\
&&p_\phi=\frac{{\rm L+sCE}}{1-s^2{\rm D}}\ ,\label{pp-gamma}
\ear
where for the $\gamma$-metric we have
\bear
&&{\rm A}=-\left(1-\frac{2 M}{r}\right)^{(2-\gamma )(\gamma -1)/2}
\left(1-\frac{M}{r}\right)^{\gamma^2-1}\frac{M\gamma}{r^3},\nn\\
&&{\rm B}=0\ ,\nn\\
&&{\rm C}=\left(1-\frac{2 M}{r}\right)^{-\gamma(\gamma +1)/2}
\left(1-\frac{M}{r}\right)^{\gamma^2-1}\frac{r-M\gamma-M}{r},\nn\\
&&{\rm D}=\left(1-\frac{2 M}{r}\right)^{(1-\gamma)\gamma-1}
\left(1-\frac{M}{r}\right)^{2(\gamma ^2-1)}\frac{M\gamma(r-M\gamma-M)}{r^4}.\nn
\ear
The radial momentum of the particle becomes
\bear\label{pr-gamma}
(p^r)^2&&=\left(1-\frac{2M}{r}\right)^{1-\gamma^2}
\left(1-\frac{M}{r}\right)^{2(\gamma^2-1)}\\
&&\times\left[p_t^2-\left(1-\frac{2M}{r}\right)^{2\gamma-1}
\frac{p_\phi^2}{r^2}-m^2\left(1-\frac{2M}{r}\right)^\gamma\right],\nonumber
\ear
and from (\ref{pr-gamma}) one can realize that the spinning
particle's motion in the non-rotating space-time is invariant under
the following simultaneous reversal of signs:
\bear {\rm L}\rightarrow-{\rm
L}, \quad s\rightarrow-s\ ,
\ear
as $p^r(s, {\rm L})=p^r(-s,-{\rm
L})$. By writing the radial momentum~(\ref{pr-gamma}) in terms
of $p_t$ and $p_\phi$, one arrives at the expression
(\ref{momentum-r2}) with the following notations:
\begin{widetext}
\bear\label{notations-gamma}
&&\alpha=\left(1-\frac{2 M}{r}\right)^{(\gamma -1) \gamma -1}
\left(1-\frac{M}{r}\right)^{2(1-\gamma ^2)} \left[-1+s^2\left(1-
\frac{2 M}{r}\right)^{(1-\gamma )\gamma-1} \left(1-\frac{M}{r}
\right)^{2(\gamma^2-1)}\frac{M\gamma (r-M\gamma-M)}{r^4}\right]^2,\nn\\
&&\beta=\left(1-\frac{2 M}{r}\right)^{-\gamma }-s^2 \left(1-\frac{2 M}{r}\right)^{-\gamma ^2-1}
\left(1-\frac{M}{r}\right)^{2(\gamma^2-1)}\frac{(r-M\gamma-M)^2 }{r^4},\nn\\
&&\delta=-s\left(1-\frac{2 M}{r}\right)^{-(\gamma -1)\gamma/2 -1}
\left(1-\frac{M}{r}\right)^{\gamma^2-1} \frac{(r-2M\gamma-M)}{r^3},\\
&&\sigma=-\left(1-\frac{2M}{r}\right)^{\gamma -1}\frac{1}{r^2}+s^2
\left(1-\frac{2 M}{r}\right)^{-(\gamma -2) \gamma -2}\left(1-\frac{M}{r}\right)^{2
\left(\gamma ^2-1\right)}\frac{M^2\gamma^2}{r^6},\nn\\
&&\rho=-m^2 \left[-1+s^2\left(1-\frac{2 M}{r}\right)^{(1-\gamma )\gamma-1}
\left(1-\frac{M}{r}\right)^{2(\gamma^2-1)}\frac{M\gamma (r-M\gamma-M)}{r^4}\right]^2,\nn
\ear
\end{widetext}
Now, by inserting the above parameters into the expression for the
effective potential $V_+$ given by (\ref{eff-potential}), one can
easily write explicitly the effective potential for the spinning
particle in the $\gamma$-metric. Given the length of the
expression for $V_+$ it is more useful to graphically study the
radial profiles for different values of the spin parameter $s$ of
the test particle in space-times corresponding to slightly oblate
(i.e. $\gamma>1$), slightly prolate (i.e. $\gamma<1$) sources. The
explicit form of $V_+$ is given in the appendix \ref{app}.

\begin{figure*}[ht]
\includegraphics[width=0.48\textwidth]{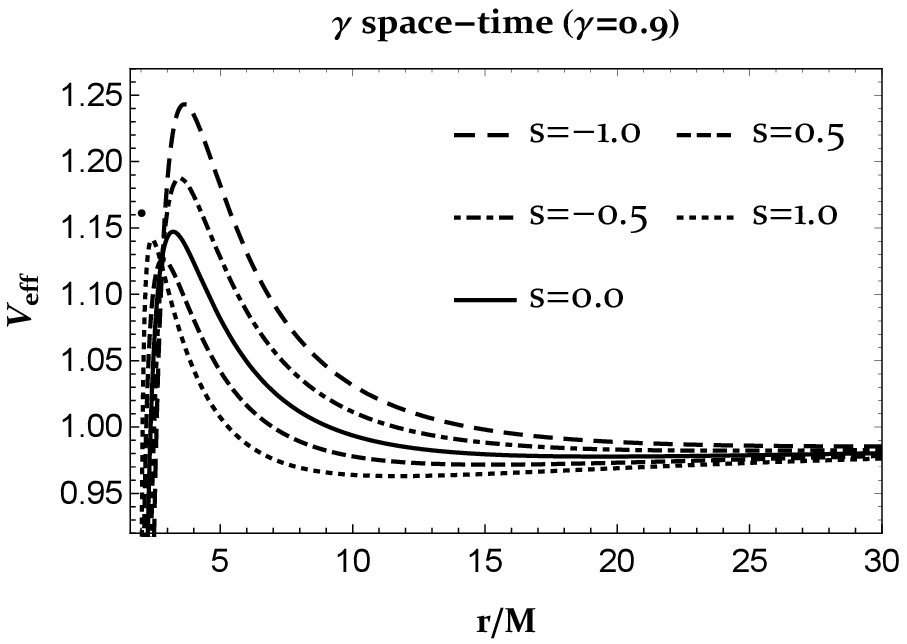}
\includegraphics[width=0.46\textwidth]{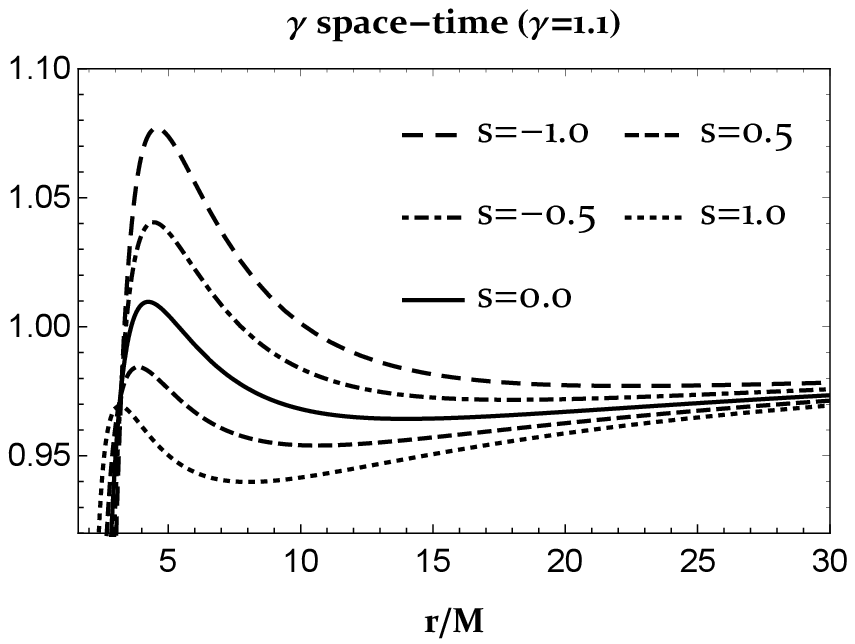}
\includegraphics[width=0.48\textwidth]{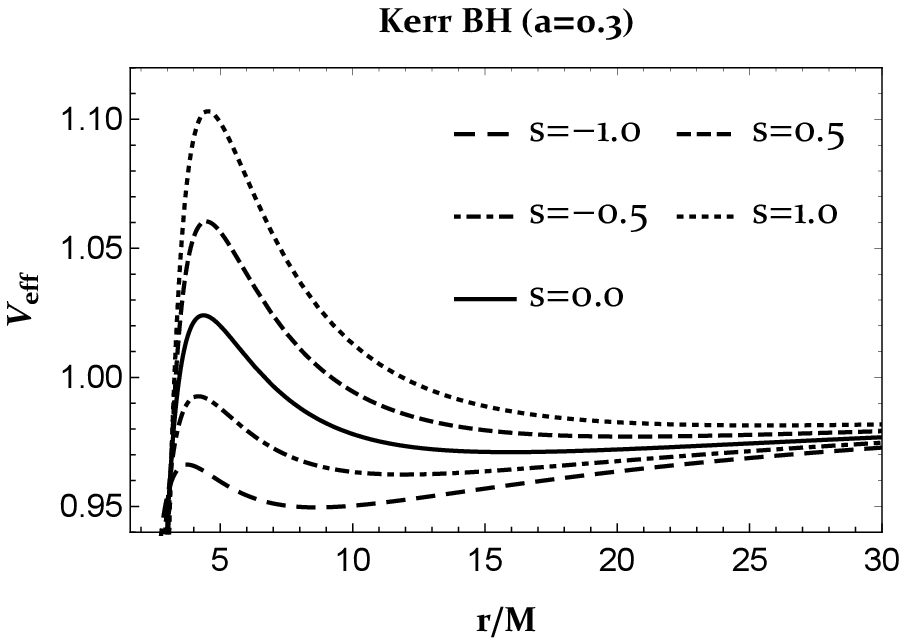}
\includegraphics[width=0.46\textwidth]{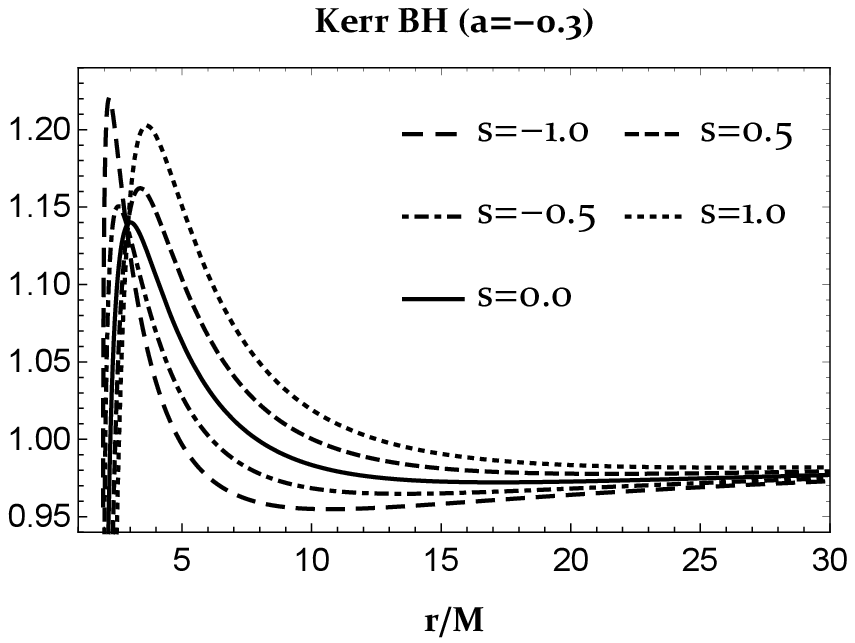}
\caption{\label{fig-veff-gamma} Top panel: radial profile of the
effective potentials of the spinning test particle in the $\gamma$
space-time with prolate ($\gamma=0.9$, left panel) and oblate
($\gamma=1.1$, right panel) source for different values of the particle's
spin. Bottom panel: radial profile of the effective potentials of
co-rotating (left panel) and counter-rotating (right panel)
spinning test particles in the Kerr black hole with different
values of spin. In all plots, the specific angular momentum of the particle is
fixed as ${\rm L/m}=4.5$.}
\end{figure*}

The effective potentials are given in Fig.~\ref{fig-veff-gamma}
and, given the possibility that the deformation parameter of a
static source may produce effects on test particles similar to
rotation of the source, we compared with the effective potentials
for the Kerr space-time.
One can see from Fig.~\ref{fig-veff-gamma} that, depending on the
values of spin and angular momentum of the particle, the effective
potential possesses two local extrema: a maximum and a minimum
that correspond to the unstable and stable circular orbits,
respectively. The effect of the spin on the value of circular
orbits can be seen from Fig.~\ref{fig-veff-gamma}. As the spin of
the particle increases, the radii of the both stable and unstable
circular orbits decrease. Conversely, as the spin of the particle
decreases, the radii of the both stable and unstable circular
orbits increase. Furthermore, similarly to what was obtained for
non-spinning particles  in~\cite{Toshmatov:PRD:2019}, the
$\gamma$-metric for an oblate spheroidal source has stable and
unstable circular orbits with bigger radii for the test particles
with respect to the $\gamma$-metric for a prolate spheroidal
source.

\begin{figure}[h]
\centering
\includegraphics[width=0.48\textwidth]{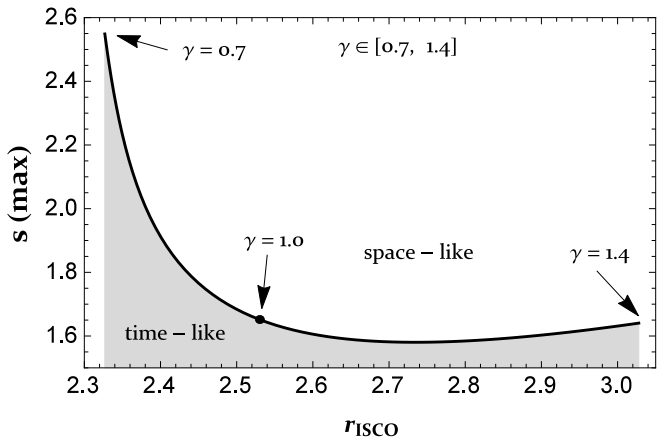}
\caption{\label{fig-super-gamma} Maximum value allowed for the
spin of test particles at the ISCO in the $\gamma$-metric as a
function of $\gamma\in[0.7, 1.4]$. The solid line separates
timelike motion, $u^\alpha u_\alpha<0$, from the spacelike,
$u^\alpha u_\alpha>0$.}
\end{figure}

It is well-known that one of the most important orbits around
compact objects in astrophysics is the ISCO, which determines the
inner edge of accretion disks. As we have seen the value of
circular orbits is affected by the spin of the test particles and
therefore the value of the ISCO is also subject to change
depending on the spin of the particles in the disk. In the case of
the $\gamma$-metric, the expression of the radial momentum of the
spinning particle~(\ref{pr-gamma}) is rather long and it can be
found in the appendix \ref{app}. Here we will restrict to
numerical results that illustrate the dependence of the ISCO on
the spin of the particles. However, before turning the attention
to circular orbits, one must remember that  for the particle's
motion to be physically realistic, we need to determine the limits
imposed by the super-luminal bound. To find the super-luminal
limit, one needs to find the kinematical 4-velocity which was
given in equations (\ref{ualpha}) for spinning particles in the
equatorial plane. The coefficients (\ref{coeffs}) in terms of the
$\gamma$ space-time become
\begin{widetext}
\bear\label{coeffs-gamma}
&&A_1=\left(1-\frac{M}{r}\right)
\left(1-\frac{2 M}{r}\right)^{\gamma -2}\frac{4M\gamma
\left[2r^2-2 (\gamma+2)M r+(\gamma +1)^2 M^2\right]}{r^5},\nonumber\\
&&B_1=-\left(1-\frac{M}{r}\right) \left(1-\frac{2
M}{r}\right)^{-\gamma -1}\frac{4M\gamma \left[r^2-(\gamma+2)M
r+\gamma(\gamma+1)M^2\right]}{r^3}, \\
&&C_1=\left(1-\frac{2 M}{r}\right)^{(1-\gamma ) \gamma}
\left(1-\frac{M}{r}\right)^{2 \gamma ^2-2}\frac{4M\gamma(r-M\gamma
-M)}{r^2}.\nonumber
\ear
\end{widetext}
Now, by inserting expressions~(\ref{Str-gamma}), (\ref{pt-gamma}),
(\ref{pp-gamma}) and (\ref{coeffs-gamma}) into
(\ref{super-generic}), one can find the super-luminal bound values
of the spinning particle in the $\gamma$ space-time. These values
for a particle located on the ISCO are presented in
Fig.~\ref{fig-super-gamma} for different values of $\gamma$.
Further, in Tab.~\ref{tab-super-gamma} we relate the values of
$s(max)$ in Fig.~\ref{fig-super-gamma} to the values of the
parameters characterizing the particle's motion at the ISCO.
\begin{table}
\caption{\label{tab-super-gamma} ISCO parameters for spinning
test particles with the super-luminal
bound ($u^\alpha u_\alpha=0$) in the $\gamma$-metric.}
\begin{ruledtabular}
\begin{tabular}{cccccc}
$\gamma$ & ${\rm s(max)}$ & ${\rm r_{ISCO}(min)}$ &  ${\rm
E_{ISCO}}$
& ${\rm L_{ISCO}}$ & ${\rm \Omega_{ISCO}}$  \\
\hline
 1.4  & 1.6405 & 3.0278 & 0.8928 & 3.4095 & 0.1318 \\
 1.3  & 1.6019 & 2.8963 & 0.8740 & 2.9298 & 0.1411 \\
 1.2  & 1.5814 & 2.7678 & 0.8511 & 2.4271 & 0.1520 \\
 1.1  & 1.5906 & 2.6444 & 0.8231 & 1.8943 & 0.1650 \\
 1.0  & 1.6518 & 2.5299 & 0.7894 & 1.3226 & 0.1809 \\
 0.9  & 1.8083 & 2.4320 & 0.7511 & 0.7093 & 0.2009 \\
 0.8  & 2.1146 & 2.3634 & 0.7155 & 0.0962 & 0.2253 \\
 0.7  & 2.5487 & 2.3269 & 0.6946 & -0.3963 & 0.2503 \\
\end{tabular}
\end{ruledtabular}
\end{table}
One can see from Tab.~\ref{tab-super-gamma} and
Fig.~\ref{fig-super-gamma} that when the space-time has prolate
deformation, the spinning particle at the ISCO is allowed to have
higher value of spin relative to the ones in the not deformed and
oblately deformed space-times. With a change of the deformation
from prolate towards oblate, the limit of spin of the particle
decreases quite rapidly till the value ${\rm s(max)}\approx1.58$
at the ISCO radius ${\rm r_{ISCO}}\approx2.74$ which correspond to
$\gamma\approx1.17$. After that value, the superluminal limit of
the spin of particle increases slowly with increasing the value of
$\gamma$.

\begin{figure*}[th]
\centering
\includegraphics[width=0.48\textwidth]{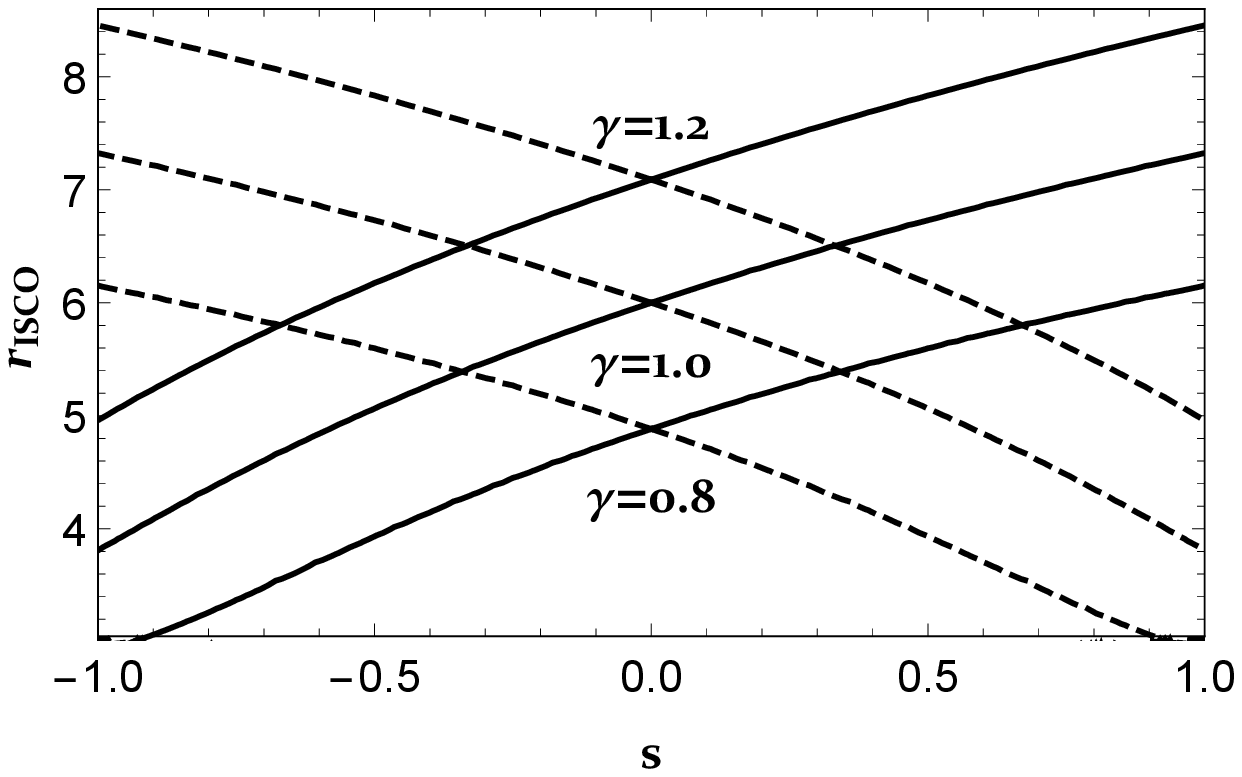}
\includegraphics[width=0.48\textwidth]{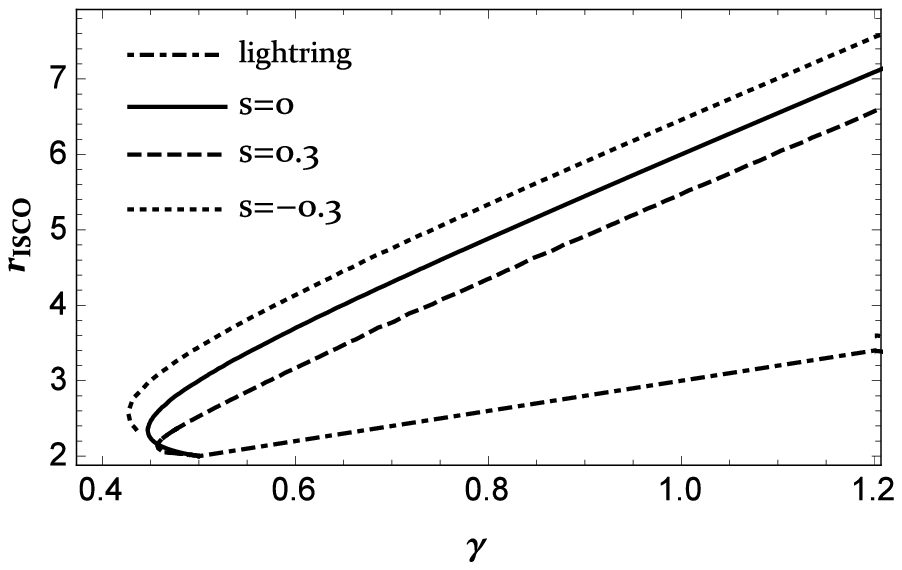}
\caption{\label{fig-isco-gamma} Left panel: Dependence of the ISCO
of spinning test particles on the equatorial plane of the
$\gamma$-metric on the particle's spin for different values of
$\gamma$. From upper to lower crossed lines correspond to the
cases of $\gamma=1.2$, $\gamma=1.0$ (Schwarzschild), and
$\gamma=0.8$, respectively. Dashed and solid lines represent ${\rm
L_{ISCO}}>0$ and ${\rm L_{ISCO}}<0$ cases, respectively. Right
panel: Dependence of the ISCO radius on the deformation parameter
$\gamma$ for different values of the particle's spin. For
completeness we illustrate also the limiting orbit give by the
photon capture radius (dot-dashed line).}
\end{figure*}

Of course, one can also calculate the ISCO for spinning particles
in the $\gamma$-metric for the values of the spin parameter that
do not exceed the super-luminal limit. In Fig.
\ref{fig-isco-gamma} we show the dependence of the ISCO radius of
the spinning test particle located on the equatorial plane in the
$\gamma$-metric on the three possible shapes: oblate (i.e.
$\gamma>1$), spherical (i.e. $\gamma=1$), and prolate (i.e.
$\gamma<1$).
\begin{table}[h]
\caption{\label{tab-isco-gamma} Characteristic parameters of ISCO
of the spinning particle moving in the $\gamma$ space-time. Where
the values corresponding to $\gamma=1.0$ represents the ones of
the Schwarzschild space-time.}
\begin{ruledtabular}
\begin{tabular}{ccccccc}
${\rm s}$ & $\gamma$ & ${\rm r_{ISCO}}$ &  ${\rm E_{ISCO}}$
& ${\rm L_{ISCO}}$ & ${\rm \Omega_{ISCO}}$ & $u^2_{{\rm ISCO}}$ \\
\hline
     & 1.2 & 7.4040 & 0.9460 & 4.0794 & 0.0767 & -0.5224 \\
     & 1.1 & 6.8600 & 0.9459 & 3.7224 & 0.0846 & -0.5208 \\
-0.2 & 1.0 & 6.3114 & 0.9457 & 3.3645 & 0.0945 & -0.5185 \\
     & 0.9 & 5.7562 & 0.9455 & 3.0052 & 0.1072 & -0.5152 \\
     & 0.8 & 5.1911 & 0.9450 & 2.6438 & 0.1245 & -0.5102 \\ \hline
     & 1.2 & 7.2501 & 0.9448 & 4.1305 & 0.0771 & -0.5150 \\
     & 1.1 & 6.7070 & 0.9446 & 3.7735 & 0.0851 & -0.5128 \\
-0.1 & 1.0 & 6.1594 & 0.9443 & 3.4156 & 0.0952 & -0.5097 \\
     & 0.9 & 5.6053 & 0.9439 & 3.0562 & 0.1081 & -0.5053 \\
     & 0.8 & 5.0415 & 0.9432 & 2.6946 & 0.1257 & -0.4986 \\ \hline
     & 1.2 & 7.0900 & 0.9436 & 4.1794 & 0.0788 & -0.5000 \\
     & 1.1 & 6.5472 & 0.9433 & 3.8223 & 0.0860 & -0.5000 \\
0.0  & 1.0 & 6.0000 & 0.9428 & 3.4641 & 0.0962 & -0.5000 \\
     & 0.9 & 5.4464 & 0.9422 & 3.1043 & 0.1095 & -0.5000 \\
     & 0.8 & 4.8832 & 0.9412 & 2.7422 & 0.1278 & -0.5000 \\ \hline
     & 1.2 & 6.9230 & 0.9422 & 4.2260 & 0.0805 & -0.4984 \\
     & 1.1 & 6.3800 & 0.9418 & 3.8685 & 0.0893 & -0.4945 \\
0.1  & 1.0 & 5.8325 & 0.9411 & 3.5097 & 0.1004 & -0.4892 \\
     & 0.9 & 5.2787 & 0.9402 & 3.1492 & 0.1150 & -0.4820 \\
     & 0.8 & 4.7153 & 0.9389 & 2.7859 & 0.1352 & -0.4715 \\ \hline
     & 1.2 & 6.7485 & 0.9408 & 4.2700 & 0.0836 & -0.4892 \\
     & 1.1 & 6.2046 & 0.9401 & 3.9118 & 0.0931 & -0.4840 \\
0.2  & 1.0 & 5.6562 & 0.9392 & 3.5521 & 0.1052 & -0.4772 \\
     & 0.9 & 5.1013 & 0.9380 & 3.1902 & 0.1213 & -0.4682 \\
     & 0.8 & 4.5368 & 0.9363 & 2.8252 & 0.1440 & -0.4554 \\
\end{tabular}
\end{ruledtabular}
\end{table}
\begin{figure*}[ht]
\includegraphics[width=0.34\textwidth]{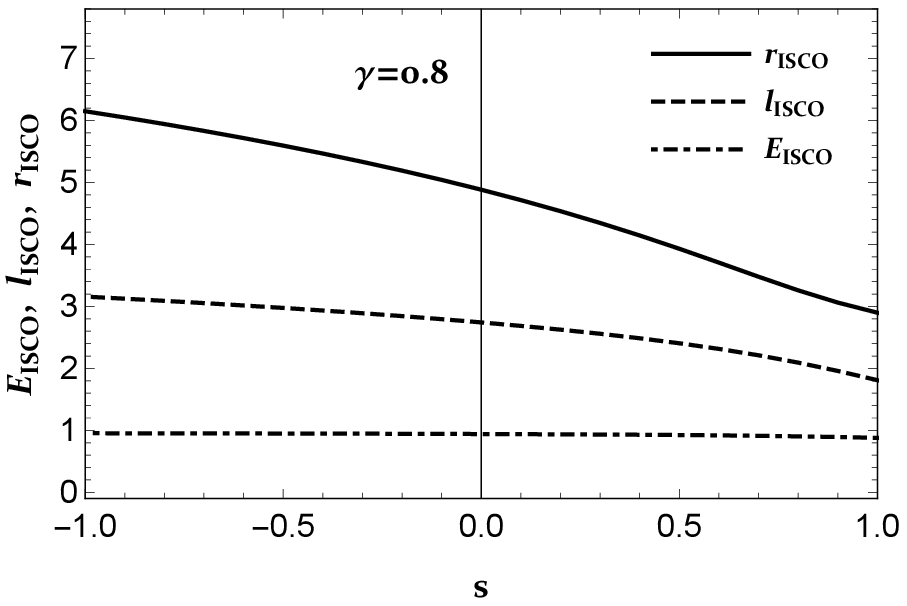}
\includegraphics[width=0.32\textwidth]{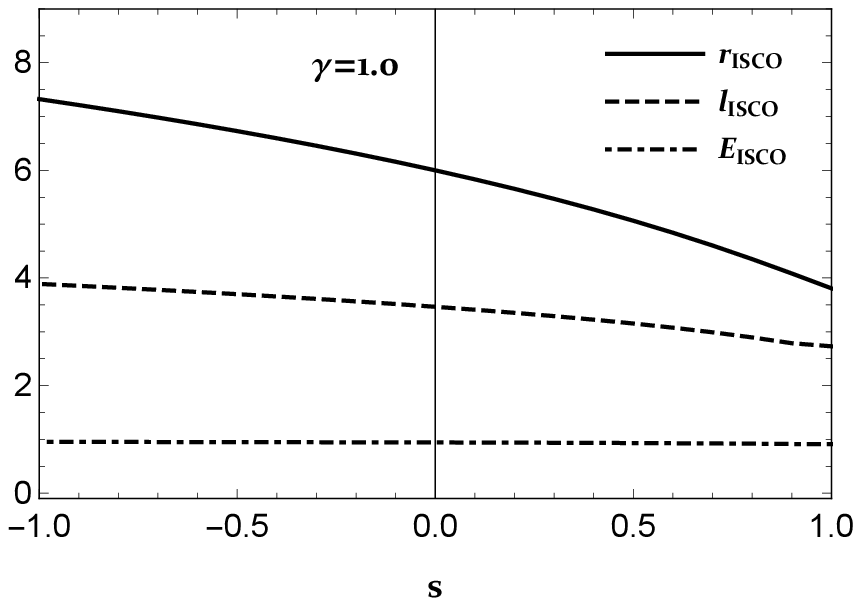}
\includegraphics[width=0.32\textwidth]{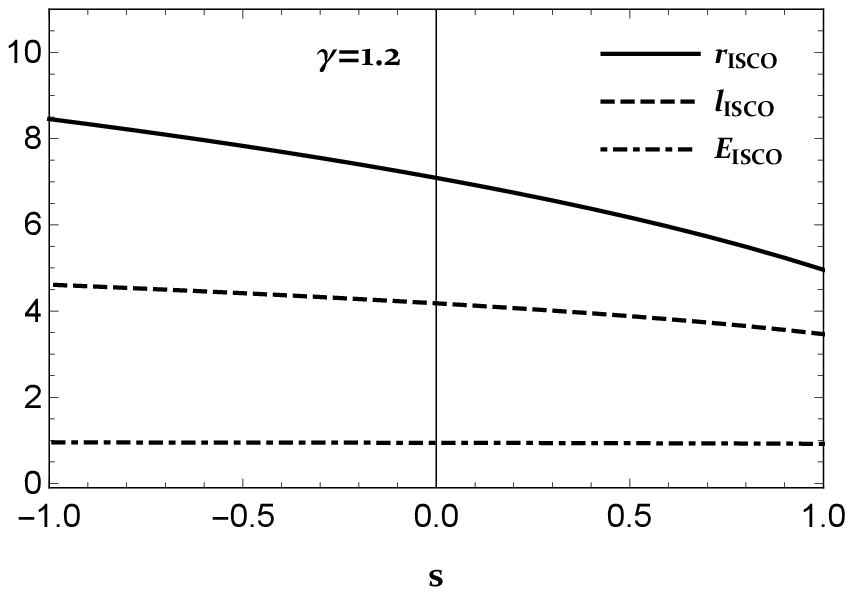}
\caption{\label{fig-elr-gamma} Spin dependence of characteristic
parameters of the test particle at the ISCO, namely energy ${\rm
E_{ISCO}}$ (dot-dashed line), total angular momentum ${\rm
l_{ISCO}=L_{ISCO}+s}$ (dashed line), and radius of ISCO ${\rm
r_{ISCO}}$ (solid line) in the $\gamma$-metric. Left panel for
prolate, central panel for spherical and right panel for oblate
sources. The intersecting vertical line corresponds to the
non-spinning particle.}
\end{figure*}
Fig.~\ref{fig-elr-gamma} shows the dependence of energy, angular
momentum and radius of the ISCO on the value of the spin
parameter, for prolate (left panel), spherical (central panel) and
oblate (right panel) geometries. One more important quantity
related to circular motion of a test particle is the particle's
angular velocity or orbital frequency. In general, the orbital
frequency of a test particle, relative to an observer at infinity
is defined by ${\rm \Omega=u^\phi/u^t}$. Then, ${\rm
\Omega_{ISCO}}$ in Tabs.~\ref{tab-super-gamma}
and~\ref{tab-isco-gamma} identifies the value of the angular
velocity of the particle at the ISCO. Finally,
Tab.~\ref{tab-isco-gamma} shows the numerical values of the above
quantities for different values of $s$ and $\gamma$. One can
easily notice that, similarly to the case of non-spinning
particles, for a fixed value of $s$, oblateness ($\gamma>1$)
implies a larger ISCO radius with respect to prolateness
($\gamma<1$).

Depending on the signs of the spin $s$ and angular momentum $L$ of
the test particles, since the value of the ISCO depends on
$s_0{\rm L_{\rm 0}}$, one can see from Fig. \ref{fig-isco-gamma}
two situations are possible, as following:
\begin{itemize}
\item[(i)] Spinning particles with spin $s_0>0$ moving
clockwise ${\rm L_{\rm 0}}>0$ have the same ISCO as the ones with
spin $s_0<0$ moving counter-clockwise ${\rm L_{\rm 0}}<0$.

\item[(ii)] Spinning particles with spin $s_0<0$ moving
clockwise ${\rm L_{\rm 0}}>0$ have the same ISCO as the ones with
spin $s_0>0$ moving counter-clockwise ${\rm L_{\rm 0}}<0$.
\end{itemize}
Also, from Fig.~\ref{fig-isco-gamma} and Tab.~\ref{tab-isco-gamma}
we see that, for fixed values of $|s|$ and $|{\rm L}|$, we have
${\rm r_{ISCO}}(s_0{\rm L_{\rm 0}}<0)>{\rm r_{ISCO}}(s_0{\rm
L_{\rm 0}}>0)$.

From Tab.~\ref{tab-isco-gamma}, and similarly to what was shown in
previous papers \cite{Toshmatov:PRD:2019,Abdikamalov:PRD:2019}, we
see that the oblate $\gamma$ space-time has larger ISCO radius
with respect to the spherical and prolate case for non spinning
test particles. However, if the test particle is spinning, the
above statement is not always correct. In some cases, depending on
the values of spin and angular momentum of the particle, the
prolate $\gamma$-metric might have a larger ISCO radius then the
corresponding oblate geometry. In particular, counter-spinning
particles (i.e. $s_0{\rm L_{\rm 0}}<0$) in the prolate geometry
may have larger ISCO than co-spinning (i.e. $s_0{\rm L_{\rm
0}}>0$) particles in the oblate geometry. This may be relevant
when it comes to the determination of the geometry around
astrophysical massive compact objects, as spinning particles in
the accretion disk may make a black hole mimicker look like a
black hole.

\section{Conclusion}\label{sec-conclusion}

In this paper we studied the motion of spinning particles in the
equatorial plane of the $\gamma$-metric, in the ``pole-dipole"
approximation by using the MPD equations with Tulczyjew-SSC. The
study of the ISCO location depending on the properties of the
particles in accretion disks is important in astrophysics as it is
the first step towards the possibility of determining the nature
of the geometry around compact objects. In our previous
paper~\cite{Toshmatov:PRD:2019} we had shown the relation of the
ISCO radius of neutral non-spinning test particles to the
deformation parameter of the $\gamma$-metric. The ISCO is bigger
than Schwarzschild's for oblate sources and smaller for prolate
sources, namely
\bear\label{riscog0}
{\rm r_{ISCO}(\gamma>1)}>{\rm
r_{ISCO}(\gamma=1)}\equiv6M>{\rm r_{ISCO}(\gamma<1)}.\nonumber\\
\ear
However, in the case of a spinning test particle, the
relation~(\ref{riscog0}) changes depending on the spin of the
particle. Therefore the most significant way to classify the
orbits is in terms of the spin-angular momentum ($s-{\rm L}$)
orientation of the particle. In Fig.~\ref{fig-isco-gamma} we
showed that
\bear {\rm r_{ISCO}}(s_0{\rm L_{\rm 0}}<0)>{\rm
r_{ISCO}}(s=0)>{\rm r_{ISCO}}(s_0{\rm L_{\rm 0}}>0)\ . \nonumber\\
\ear
As it was mentioned above, the kinematical 4-velocity and
dynamical 4-momentum of the spinning particle are not always
parallel. Therefore, despite the fact that the normalization of
the 4-momentum always hold, the kinematical 4-velocity may exceed
the speed of light, which is not physical. Therefore, one must
impose an extra condition to ensure that the particle's motion is
always time-like. As a consequence of this we have shown that the
allowed spin of particles located on the ISCO of $\gamma$-metric
can be higher for prolate sources with respect to oblate sources.

The final aim is to compare the theoretical predictions for
accretion disks around black holes and black hole mimickers. In
this view, it is important to compare the results obtained for the
$\gamma$-metric with the corresponding situation in the Kerr
geometry.
In~\cite{Suzuki:PRD:1998,Semerak:MNRAS:1999,Semerak:MNRAS:2007,Plyatsko:PRD:2013}
it was shown that the value of the ISCO for spinning particles in
the Kerr space-time has a wider range with respect to the case of
the spinless particles. However, the lower limit of the ISCO, i.e.
${\rm r_{ISCO}}\geq M$, remains unchanged by the introduction of
the spin of test particles. Similarly, here we have shown that the
value of the ISCO for spinning particles in the $\gamma$-metric
also has a wider range with respect to the case of the spinless
particles. However, the lower limit of the value of the ISCO
radius must remain larger than singular surface, ${\rm
r_{ISCO}}>2M$, for small departures from spherical symmetry.
Therefore, if the mass of the compact object is measured through a
different method, the observation of an ISCO radius ${\rm
r_{ISCO}}>2M$ alone would not allow to determine if the central
object is described by the Kerr geometry or a black hole mimicker
with non vanishing quadrupole moment.

\section*{Acknowledgments}

The work was developed under the Nazarbayev University Faculty
Development Competitive Research Grant No.~090118FD5348. The
authors acknowledge the support of the Ministry of Education of
Kazakhstan's target program IRN:~BR05236454 and Uzbekistan
Ministry for Innovation Development Grants No.~VA-FA-F-2-008 and
No.~YFA-Ftech-2018-8.

\label{lastpage}

\bibliography{gamma_references}

\appendix
\section{Effective potential}\label{app}
For completeness, we present here the complete analytical
expression for the effective potential of spinning particles in
the $\gamma$-metric
\begin{widetext}
\bear V_+&&=\left(1-\frac{2 M}{r}\right)^{(2-\gamma)(\gamma -1)/2}
\left(1-\frac{M}{r}\right)^{\gamma ^2-1}\frac{s r[r-(2 \gamma +1)
M]}{r^4-s^2 z}\nonumber\\
&&+\left(1-\frac{2 M}{r}\right)^{\gamma/2}\frac{r^4+s^2
w}{r^4-s^2z}\sqrt{\left(1-\frac{2
M}{r}\right)^{\gamma-1}\frac{L^2}{r^2}+m^2\left(1-\frac{s^2z}{r^4}\right)},
\ear
\end{widetext}
where
\begin{widetext}
\bear &&w=\left(1-\frac{M}{r}\right)^{2 \left(\gamma
^2-1\right)}\left(1-\frac{2 M}{r}\right)^{-\gamma ^2+\gamma -1}
M\gamma(\gamma M+M-r),\nonumber\\
&&z=\left(1-\frac{M}{r}\right)^{2 \left(\gamma
^2-1\right)}\left(1-\frac{2
M}{r}\right)^{-\gamma^2+\gamma-1}(\gamma M+M-r)^2.\nonumber
\ear
\end{widetext}

\end{document}